\RequirePackage{ifpdf}
\ifpdf
\else
\PassOptionsToPackage{dvipdfmx}{graphicx}
\PassOptionsToPackage{dvipdfmx}{hyperref}
\fi

\documentclass[12pt,a4paper]{article}
\usepackage{jheppub}

\allowdisplaybreaks[4]
\usepackage{amsmath,bm}
\usepackage{subfigure}

\title{
Energy extraction from AdS black holes via superradiance
}

\author[1]{Takaaki Ishii}
\author[2]{Youka Kaku}
\author[3]{Keiju Murata}
\affiliation[1]{Department of Physics, Rikkyo University, Toshima, Tokyo 171-8501, Japan}
\affiliation[2]{Department of Physics, Nagoya University, Chikusa, Nagoya 464-8602, Japan}
\affiliation[3]{Department of Physics, College of Humanities and Sciences, Nihon University, Sakurajosui, Tokyo 156-8550, Japan}
\emailAdd{ishiitk@rikkyo.ac.jp}
\emailAdd{kaku.yuka.g4@s.mail.nagoya-u.ac.jp}
\emailAdd{murata.keiju@nihon-u.ac.jp}

\abstract{%
Superradiance is known as a wave amplification process caused by rotating or charged black holes. 
We argue that the superradiance of stationary black holes in asymptotically AdS spacetimes can be characterized by the ability of energy extraction. 
Specifically, we demonstrate that energy can be extracted from Reissner-Nordstr\"{o}m-AdS$_4$ and Kerr-AdS$_4$ under appropriate time-dependent boundary conditions at conformal boundaries.
This indicates that energy can be extracted from thermal states dual to these black holes by applying appropriate time-dependent sources.
We also show that the energy extraction can be realized as a reversible process. 
}

\preprint{RUP-22-14}

\begin{document}
\maketitle

\section{Introduction}

Black holes can exhibit wave amplification processes called superradiance (see Ref.~\cite{Brito:2015oca} for a comprehensive review of black hole superradiance). In the first place, this phenomenon is of interest in four-dimensional asymptotically flat black holes, which are characterized by three conserved charges: mass, electric charge, and angular momentum. Let us focus on a Kerr black hole. Radiation in a certain frequency band around the Kerr black hole is amplified by a superradiant scattering, and mass and angular momentum are extracted from the black hole. This implies that the black hole loses a fraction of its hairs, while the area increases because of Hawking's area theorem. This phenomenon for Kerr black holes is called rotational superradiance. Similar wave amplification and extraction of mass and electric charges can occur for Reissner-Nordstr\"{o}m black holes if electrically charged radiation is considered, and it is called charged superradiance. Kerr-Newman black holes are subject to both.

If there is a potential barrier outside the horizon like in asymptotically AdS spacetimes, the superradiance is related to a more dramatic consequence that black hole spacetime can be unstable. The wave amplified by the superradiance is reflected by the potential barrier. The wave is then scattered by the black hole and amplified again by the superradiance. This leads to instability called the superradiant instability.
Recent prosperity of AdS spacetimes follows the historic claiming of the AdS/CFT duality \cite{Maldacena:1997re}, and for rotating AdS black holes, rotational superradiant instability has been widely studied \cite{Hawking:1999dp,Cardoso:2004hs,Kunduri:2006qa,Cardoso:2006wa,Murata:2008xr,Kodama:2009rq,Dias:2011at,Dias:2013sdc,Cardoso:2013pza}. 
The outcome of the instability has been found to be deformed rotating black holes with fewer symmetries \cite{Dias:2011at,Dias:2015rxy,Ishii:2018oms,Ishii:2019wfs,Ishii:2021xmn}. Their dual field theory interpretation is nevertheless unclear since the rotational superradiant instability is seen only for small AdS black holes.
For charged AdS black holes, the instability leading to the condensation of a charged scalar field has been noticed \cite{Gubser:2008px} and suggested to correspond to superconductivity or superfluidity in the dual field theory \cite{Hartnoll:2008vx,Hartnoll:2008kx}. We will discuss the relation of this instability with the charged superradiance. The charged instability is of more interest to the dual field theory application compared with the rotational one because the former can be observed for large black holes.

As described above, when the superradiance is discussed in asymptotically AdS spacetimes, the instability has been mainly focused on.  
How can we characterize the superradiance for stable charged/rotating black holes in AdS? How can we observe the superradiance in the dual field theory?
We will address these questions in this paper.
We study superradiance in asymptotically AdS spacetimes by applying a source with a monochromatic frequency from the AdS boundary. 
We use two setups for demonstrations: 1) the perturbation of a four-dimensional Reissner-Nordstr\"{o}m-AdS$_4$ black hole (RNAdS$_4$) with a flat horizon topology by a charged scalar field, and 2) that of a Kerr-AdS black hole by a neutral scalar field. 
It is shown that a monochromatic source with a frequency in a certain frequency band does negative work to the AdS bulk, implying the extraction of energy from the black holes.
We hence characterize the superradiance of the AdS black holes by the ability of the energy extraction.
Periodic driving of asymptotically AdS spacetimes has been considered mainly motivated by quench processes and thermalization in holography \cite{Auzzi:2012ca,Li:2013fhw,Natsuume:2013lfa,Auzzi:2013pca,Rangamani:2015sha,Hashimoto:2016ize,Kinoshita:2017uch,Biasi:2017kkn,Ishii:2018ucz,Biasi:2019eap}. Here we shed light on the aspect of the superradiance and energy extraction for the periodic driving.

Here, we comment on the charged superradiant instability in AdS. It may have been sometimes said that the superradiant instability is for small black holes because large rotating AdS black holes are superradiant stable. Meanwhile, the instability of a charged scalar field for large charged black holes is often associated with AdS near horizon instability, whereas that for small black holes is called superradiant instability~\cite{Brito:2015oca,Dias:2018zjg}. However, if we characterize the superradiance by the ability of the energy extraction from a stationary black hole, the RNAdS$_4$ with the flat horizon also can be considered to exhibit the superradiance.

This paper is organized as follows. In section~\ref{sec:rn}, we consider charged superradiance for a probe charged scalar field in RNAdS$_4$ with asymptotically Poincar\'{e} AdS boundary. In section~\ref{sec:kerr}, rotational superradiance is discussed for a scalar field around a Kerr-AdS black hole. The conclusion and discussion are given in section \ref{sec:sum}. Appendices contain technical details omitted in the main text.

\section{Charged superradiance in AdS}
\label{sec:rn}

In this section, we consider the perturbation of the RNAdS$_4$ by a charged scalar field. Throughout this paper, we use units in which the AdS radius is unity.

\subsection{Reissner-Nordstr\"{o}m-AdS$_4$ spacetime}

As a background solution that admits superradiance, we consider the RNAdS$_4$ with the planar horizon:
\begin{equation}
 ds^2=\frac{1}{z^2}\left[-F(z)dt^2+\frac{dz^2}{F(z)}+dx^2+dy^2\right]\ ,\quad F(z)=1-2Mz^3+\frac{1}{4}Q^2z^4\ .
\label{RNAdS4}
\end{equation}
The horizon is located at $z=z_+$ given by the smallest positive root of $F(z)=0$. The Maxwell field is given by 
\begin{equation}
 A=A_t dt\ ,\quad A_t=-Q z\ ,
\end{equation}
where we set $A_t|_{z=0}=0$ using the $U(1)$-gauge freedom.

The black hole is equipped with thermodynamic quantities.
The parameters $M$ and $Q$ are proportional to the mass $\mathcal{M}$ and electric charge $\mathcal{Q}$ densities of the black hole as
\begin{equation}
 \mathcal{M}=\frac{2M}{\kappa^2} \ ,\quad \mathcal{Q}=\frac{Q}{2\kappa^2}\ ,
\label{MQdef}
\end{equation}
where the constant $\kappa^2$ is related to Newton's constant of gravitation $G$ as $\kappa^2=8\pi G$. 
The entropy density $\mathcal{S}$, Hawking temperature $T$ and electric potential on the horizon $\Phi_H=-A_t|_{z=z_+}$
are given by
\begin{equation}
 \mathcal{S}=\frac{2\pi}{\kappa^2 z_+^2}\ ,\quad
T=
\frac{12-z_+^4 Q^2}{16\pi z_+}\ ,\qquad
 \Phi_H=Qz_+\ .
\label{Setc}
\end{equation}
In our gauge, $\Phi_H$ corresponds to the chemical potential of the $U(1)$.
For regularity of the spacetime, we require $T\geq 0$, giving $Q^2\leq 12/z_+^4$.
The thermodynamic quantities satisfy the first law of black hole mechanics,
\begin{equation}
 d\mathcal{M} = Td\mathcal{S} + \Phi_H d\mathcal{Q}\ .
 \label{firstlaw}
\end{equation}

\subsection{Charged scalar field perturbation}
\label{CSP}

We perturb the RNAdS$_4$ background by a charged scalar field with a negative mass squared $\mu^2=-2$. The perturbation equation is given by
\begin{equation}
 D^2\Psi+2\Psi=0\ ,
\label{KG}
\end{equation}
where $D_\mu \Psi = (\partial_\mu - ie A_\mu)\Psi$ is the gauge covariant derivative and $e$ denotes the $U(1)$-charge of the scalar field.

The Klein-Gordon equation \eqref{KG} can be brought into a form convenient for our analysis.
Writing the scalar field as $\Psi(t,x,y,z)=z\psi(z)e^{-i\omega t + i\bm{k}\cdot \bm{x}}$, we obtain a Schr\"{o}dinger-like equation,
\begin{align}
 &\left[-\frac{d^2}{dr_\ast^2} + V(z)\right]\psi = (\omega + eA_t)^2 \psi\ ,\label{SchEq}\\
 &V(r)=\frac{F}{z^2} (k^2 z^2 - zF' + 2 F -2)\ ,
\end{align}
where $r_\ast = -\int_0^z dz'/F(z')$ is the tortoise coordinate. The AdS boundary $z=0$ and horizon $z=z_+$ correspond to $r_\ast = 0$ and $r_\ast = -\infty$, respectively.
The asymptotic solution to the above equation near the AdS boundary takes the form
\begin{equation}
\psi(z)=\psi_1  + \psi_2 z + \cdots \ ,
\label{psiexpand}
\end{equation}
where $\psi_1$ and $\psi_2$ are unfixed in the boundary analysis.
Near the horizon, we impose the ingoing wave boundary condition as
\begin{equation}
 \psi(z)\simeq C e^{-i(\omega-e\Phi_H) r_\ast}\ ,
\label{ingoing}
\end{equation}
where $C$ is a complex constant that is free in the near horizon analysis.

A relation among the three coefficients $\psi_1,\psi_2, C$ can be obtained through the Wronskian.
Let us define the Wronskian as
\begin{equation}
 W=\psi^\ast \frac{d\psi}{dr_\ast} - \psi \frac{d\psi^\ast}{dr_\ast}\ .
\label{Wronsdef}
\end{equation}
Upon using Eq.(\ref{SchEq}), it can be shown that the Wronskian is conserved along the $r_\ast$-direction: $dW/dr_\ast = 0$. 
The Wronskian can be evaluated at the horizon and AdS boundary by plugging Eqs.(\ref{psiexpand}) and (\ref{ingoing}), respectively, to Eq.(\ref{Wronsdef}). Equating the results, we obtain
\begin{equation}
  \psi_1\psi_2^\ast-\psi_1^\ast\psi_2=-2i(\omega-e\Phi_H)|C|^2\ .
\label{Wrons}
\end{equation} 
In the following, 
we will show that this quantity is related to the work done by the scalar field source on the AdS boundary.

\subsection{Energy extraction through superradiance}
\label{energyextract}

The asymptotic behavior of the scalar field $\Psi$ near the AdS boundary is related to the expectation value of the scalar operator and its source in the dual field theory.
From Eq.(\ref{KG}), the scalar field has the series solution near the AdS boundary of the form
\begin{equation}
 \Psi(t,x,y,z)=\Psi_1(t,x,y) z + \Psi_2(t,x,y) z^2 + \cdots \ .
\label{phiexpand}
\end{equation}
From the AdS/CFT dictionary~\cite{Gubser:1998bc,Witten:1998qj}, 
the leading term in the expansion is interpreted as the source $J\equiv \Psi_1$ that couples to the scalar operator $\mathcal{O}$ in the boundary theory. 
The expectation value of the latter is given by the sub-leading term in Eq.\eqref{phiexpand}, $\langle \mathcal{O} \rangle = \Psi_2$.
(It could depend on the counterterm of the on-shell action in general, but it does not in the present case. See appendix~\ref{sec:WTI} for details.) 

The source $J$, which has the $U(1)$-charge $e$, can be considered as the ``source'' of the $U(1)$ current in the boundary theory. 
In the presence of $J$, the divergence of the electric current is given by
\begin{equation}
 \nabla_i \langle j^i \rangle = ie ( \langle \mathcal{O}\rangle^\ast J - \langle \mathcal{O}\rangle  J^\ast)\ ,
\label{dj}
\end{equation}
where $\nabla_i$ represents the covariant derivative with respect to the boundary metric $\gamma_{ij}=\lim_{\epsilon\to 0} z^2 h_{ij}|_{z=\epsilon}$ 
where $h_{ij}$ is the induced metric of $z=\epsilon$. (For the RNAdS$_4$~(\ref{RNAdS4}), we have $\gamma_{ij}=\eta_{ij}$.)
This formula follows from the $U(1)$-gauge invariance of the bulk on-shell action. For the derivation, see  appendix~\ref{sec:WTI} or Ref~\cite{Skenderis:2002wp}.

Also, the source $J$ does work in the boundary theory, changing the energy.
The divergence of the energy momentum tensor of the boundary theory satisfies 
\begin{equation}
 \nabla^j \langle T_{ij} \rangle  = F_{ij} \langle j^j \rangle   
+ \langle \mathcal{O}\rangle^\ast D_i J + \langle \mathcal{O}\rangle D_i J^\ast \ ,
\label{dT}
\end{equation}
where $F_{ij}=\partial_i A_j - \partial_j A_i$ and $A_i$ is the Maxwell field at the AdS boundary. The boundary gauge covariant derivative is defined by 
$D_iJ=(\partial_i-ieA_i)J$.
Eq.(\ref{dT}) follows from the diffeomorphism invariance of the bulk on-shell action. See again appendix~\ref{sec:WTI} or Ref~\cite{Skenderis:2002wp}.
Eqs.(\ref{dj}) and (\ref{dT}) are known as the Ward-Takahashi identities.

The source induces changes of $\mathcal{M}$ and $\mathcal{Q}$ because of the Ward-Takahashi identities.
Let us take two constant time slices $\Sigma_1$ and $\Sigma_2$ at times $t=t_1$ and $t=t_2$ ($t_2>t_1$). 
The difference of the electric charges on $\Sigma_1$ and $\Sigma_2$ is then given by
\begin{equation}
\begin{split}
  \Delta \mathcal{Q}V  &= -\int d\Sigma_2 \langle j_{i} \rangle n^i + \int d\Sigma_1 \langle j_{i} \rangle n^i
=\int d^3x \sqrt{-\gamma}  \nabla^i \langle j_{i} \rangle \\
&=ie \int_{t=t_1}^{t=t_2} d^3x \sqrt{-\gamma}   ( \langle \mathcal{O}\rangle^\ast J - \langle \mathcal{O}\rangle  J^\ast)\ ,
\end{split}
\end{equation} 
where $V$ is the infinite volume of the $t=$constant surface and $n^i$ is the unit normal of $\Sigma_1$ and $\Sigma_2$. 
(Note that $\mathcal{M}$ and $\mathcal{Q}$ are the mass and charge ``densities'' in Eq.(\ref{MQdef}).)
In the second equality, the divergence theorem is used. (Note that the minus sign is needed for the surface integral when the boundary is the spacelike hypersurface~\cite{Poisson:2009pwt}.) The last equality follows from Eq.(\ref{dj}). 
This equation implies that the change of the charge per unit time and volume is given by
\begin{equation}
  \dot{\mathcal{Q}}=ie( \langle \mathcal{O}\rangle^\ast J - \langle \mathcal{O}\rangle  J^\ast)\ ,
\label{dotQ0}
\end{equation}
Similarly, we obtain the difference of the energy between $\Sigma_1$ and $\Sigma_2$ as 
\begin{equation}
\begin{split}
  \Delta \mathcal{M}V  &= \int d\Sigma_2 T_{ij}k^i n^j - \int d\Sigma_1 T_{ij}k^i n^j\\
&=-\int d^3x \sqrt{-\gamma}  \nabla^j (T_{ij}k^i)
= -\int d^3x \sqrt{-\gamma}  (\nabla^j T_{ij}) k^i\\
&= -\int d^3x \sqrt{-\gamma} \left[F_{ij} \langle j^j \rangle   
+ \langle \mathcal{O}\rangle^\ast D_i J + \langle \mathcal{O}\rangle D_i J^\ast \right]k^i \ ,
\end{split}
\label{Mchange}
\end{equation} 
where $k=\partial_t$ is the timelike Killing vector.
In the third equality, we used the Killing equation $\nabla_i k_j + \nabla_j k_i=0$. 
Eq.(\ref{dT}) is used in the last equality.
Because we adopt the gauge in which the asymptotic value of the Maxwell field is zero as well as $F_{ij}=0$ for the RNAdS$_4$, the change of the energy per unit time and unit volume is given by
\begin{equation}
\dot{\mathcal{M}}= -( \langle \mathcal{O}\rangle^\ast \dot{J} - \langle \mathcal{O}\rangle \dot{J}^\ast )\ .
\label{dotM0}
\end{equation}

The formulae \eqref{dotQ0} and \eqref{dotM0} can then be translated to the conditions of superradiance.
Specifically for the source with a monochromatic frequency $\omega$ and a single wave number $\bm{k}$,
\begin{equation}
 J(t,x,y)=\psi_1 e^{-i\omega t + i\bm{k}\cdot \bm{x}}\ ,\quad
 \langle \mathcal{O}(t,x,y)\rangle =\psi_2 e^{-i\omega t + i\bm{k}\cdot \bm{x}}\ ,
\end{equation}
we find
\begin{align}
&\dot{\mathcal{Q}}=ie (\psi_1 \psi_2^\ast -\psi_1^\ast \psi_2) =2e (\omega-e\Phi_H) |C|^2\ ,\label{dotQ}\\
&\dot{\mathcal{M}}= i\omega (\psi_1 \psi_2^\ast-\psi_1^\ast \psi_2)=2\omega(\omega-e\Phi_H) |C|^2 \label{dotM}\ ,
\end{align}
where the conservation of the Wronskian~(\ref{Wrons}) is used.  
In appendix~\ref{sec:BReac}, we provide an alternative derivation of Eqs.(\ref{dotQ}) and (\ref{dotM}) by taking into account the backreaction of the scalar field to the metric and Maxwell field. 

We find that $\dot{\mathcal{M}}$ becomes negative for
\begin{equation}
 0<\omega<e\Phi_H\quad(e\Phi_H>0)\ ,\qquad e\Phi_H < \omega <0\quad(e\Phi_H<0)\ .
\label{SRcond}
\end{equation}
The source does negative work for the frequencies satisfying the above conditions. 
In other words, we can extract energy from the thermal state dual to RNAdS$_4$ by applying a monochromatic source with its frequency in the range \eqref{SRcond}.
In the case of the asymptotically flat charged black holes, the inequality~(\ref{SRcond}) appears as the superradiant condition, i.e., an incident wave towards the black hole is reflected as waves with larger amplitudes. For asymptotically AdS spacetimes, this inequality gives the condition for the energy extraction through the external source.

Note that the first law of black hole mechanics \eqref{firstlaw} implies the change of the entropy density to be
\begin{equation}
 T\dot{\mathcal{S}}=\dot{\mathcal{M}}-\Phi_H \dot{\mathcal{Q}}=2(\omega-e\Phi_H)^2 |C|^2\ .
\label{dotS}
\end{equation}
This is always positive, consistent with Hawking's area theorem. 

Eqs.(\ref{dotQ}), (\ref{dotM}) and (\ref{dotS}) indicate that we can extract energy from the black hole by a reversible process.
Writing the frequency as $\omega=e\Phi_H - \delta \omega$ ($\delta \omega \ll e\Phi_H$), we find 
$\dot{\mathcal{M}}=\mathcal{O}(\delta\omega)$, $\dot{\mathcal{Q}}=\mathcal{O}(\delta\omega)$ and $\dot{\mathcal{S}}=\mathcal{O}(\delta\omega^2)$.
Hence, when $\omega$ is sufficiently close to $e\Phi_H$, the mass and charge can be changed while the entropy is fixed. The energy extraction process can be isentropic (adiabatic and reversible).

\begin{figure}[t]
\centering
\includegraphics[scale=0.7]{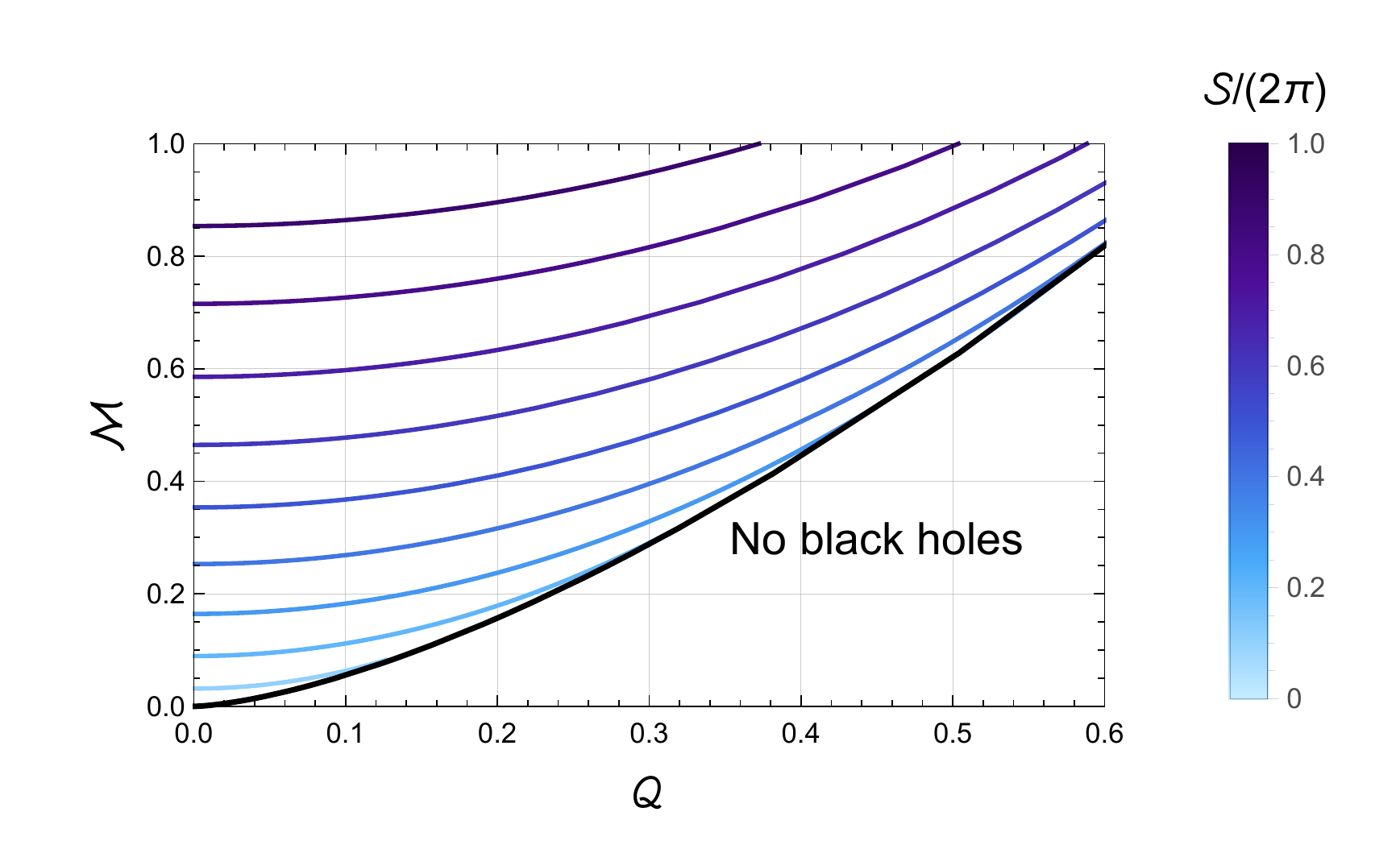}
\caption{Contours of the entropy density of the RNAdS$_4$ for $S/2\pi=0.1,0.2,\dots,0.9$.}
\label{fig:MQ}
\end{figure}

From Eqs.(\ref{MQdef}) and (\ref{Setc}), the mass density can be given as a function{} of the charge and entropy densities as
\begin{equation}
 \mathcal{M}(\mathcal{S},\mathcal{Q})=\left(\frac{\mathcal{S}}{2\pi}\right)^{3/2}\left\{1+\mathcal{Q}^2\left(\frac{\mathcal{S}}{2\pi}\right)^{-2}\right\}\ ,
\end{equation}
where we set $\kappa^2=1$ for simplicity. The $\kappa$-dependence can be easily recovered by replacing $\mathcal{M}\to\kappa^2 \mathcal{M}$, $\mathcal{S}\to\kappa^2 \mathcal{S}$, $\mathcal{Q}\to\kappa^2 \mathcal{Q}$. The condition for the absence of the naked singularity is written as $\mathcal{M}\geq 4\cdot 3^{-3/4} \mathcal{Q}^{3/2}$, where the equality is satisfied for the extreme RNAdS$_4$. 
Contours of the entropy density in the $(\mathcal{M},\mathcal{Q})$-space are shown in Fig.~\ref{fig:MQ}.\footnote{If dimensionless variables are introduced as $\widetilde{\mathcal{M}} = \mathcal{M}\mathcal{S}^{-3/2}$ and $\widetilde{\mathcal{Q}} = \mathcal{Q}\mathcal{S}^{-2}$, these contours merge into a single curve. This is because of the planar horizon we consider for simplicity. However, generalization to non-planar horizons such as asymptotically global AdS is straightforward, and hence we discuss isentropic contours in the $(\mathcal{M},\mathcal{Q})$-space rather than reducing them to the aforementioned dimensionless variables.}
In an isentropic process, $\mathcal{M}$ and $\mathcal{Q}$ change along a contour. In a non-isentropic process, the trajectory of the change in $(\mathcal{M},\mathcal{Q})$-space goes upward from the contour of the initial entropy.

How much energy can we maximally extract from the black hole? 
Let the initial mass and charge densities be $\mathcal{M}_\textrm{ini}$ and $\mathcal{Q}_\textrm{ini}$. The initial entropy $\mathcal{S}_\textrm{ini}$ is determined by solving 
\begin{equation}
 \mathcal{M}_\textrm{ini} = \mathcal{M}(\mathcal{S}_\textrm{ini},\mathcal{Q}_\textrm{ini})\ .
\label{Mini}
\end{equation}
As long as the black hole has nonzero charges, we can extract energy using the frequency satisfying Eq.(\ref{SRcond}).
The energy extraction terminates when all electric charges are extracted $\mathcal{Q}=0$ (i.e.~when the trajectory in Fig.~\ref{fig:MQ} reaches the vertical axis).  
The final mass is minimal in the isentropic process because the trajectory of the change for general processes is in the upper side of the isentropic contour for $\mathcal{S}_\textrm{ini}$ in Fig.~\ref{fig:MQ}. 
It follows that the maximum energy extraction is for the isentropic process, where the final mass of the black hole is given by 
$\mathcal{M}_\textrm{fin}=\mathcal{M}(\mathcal{S}_\textrm{ini},0)$. Therefore, 
the maximum energy per unit volume that we can extract from the black hole is given by a function of $\mathcal{M}_\textrm{ini}$ and $\mathcal{Q}_\textrm{ini}$ as
\begin{equation}
 \Delta \mathcal{M}=\mathcal{M}_\textrm{ini}-\mathcal{M}_\textrm{fin}=\mathcal{Q}_\textrm{ini}^2\left(\frac{\mathcal{S}_\textrm{ini}}{2\pi}\right)^{-1/2}\ ,
\end{equation}
where $\mathcal{S}_\textrm{ini}(\mathcal{M}_\textrm{ini},\mathcal{Q}_\textrm{ini})$ is determined by solving Eq.(\ref{Mini}).

\subsection{Detailed calculation of the energy extraction}

In this subsection, we numerically calculate the variation of the mass density $\dot{\mathcal{M}}$ for different wave frequencies $\omega$ and black holes charge densities $Q$. We will show that the relation \eqref{dotM} is satisfied indeed. 
Since there is a scaling symmetry for the flat horizon spacetime, we can set $z_+=1$. Besides, the overall factor of the scalar field is free because it is a linear perturbation. Here, we assume that we apply an external source with constant amplitude in the dual picture in QFT, and we evaluate the variation of the mass density normalized by the magnitude of the boundary source: $\dot{\mathcal{M}}/|\psi_1|^2$. 
The electric charge is fixed to $e=1$ and the spatial wave number is fixed to $\bm{k}=0$, but the behavior is qualitatively the same for other $e$, $\bm{k}$. 

It is known that the RNAdS$_4$ becomes unstable against the charged scalar perturbation in a certain parameter space even without the source~\cite{Gubser:2008px,Hartnoll:2008kx}. 
At the onset of the instability, the response rescaled by the source $\langle \mathcal{O} \rangle/\psi_1$ hence diverges.
The instability of the RNAdS$_4$ is discussed together with quasinormal modes in appendix~\ref{sec:qnm}.

In Fig.~\ref{fig:dM_RN}, we show the $\omega$-dependence of the rescaled variation of the mass density $\dot{\mathcal{M}}/|\psi_1|^2$ for $Q=2.0,\,2.1,\,\cdots,2.9$. 
Since the horizontal axis is rescaled as $\omega/(e\Phi_H)$, the superradiant condition is satisfied for $0<\omega/(e\Phi_H)<1$, which is shown as the red region. It is obvious that $\dot{\mathcal{M}}$ is negative  when the superradiant condition is satisfied, and we can extract energy from the charged black hole. Also, as the charge is increased, a sharp peak appears at $\omega/(e\Phi_H)\lesssim 1$. 
In other words, we can extract more energy from a black hole per unit time per an external source magnitude by applying the scalar field with frequency $\omega/(e\Phi_H)\lesssim 1$ just before the onset of instability. At the onset of instability $Q=2.981$, the peak diverges since our definition of the rescaled mass variation is singular at that time. Beyond this $Q$, the RNAdS$_4$ becomes unstable and is replaced with black holes with scalar hair \cite{Hartnoll:2008kx}. Hence, here we do not include $Q$ bigger than the onset.

\begin{figure}[t]
    \centering
    \includegraphics[width=0.8\linewidth]{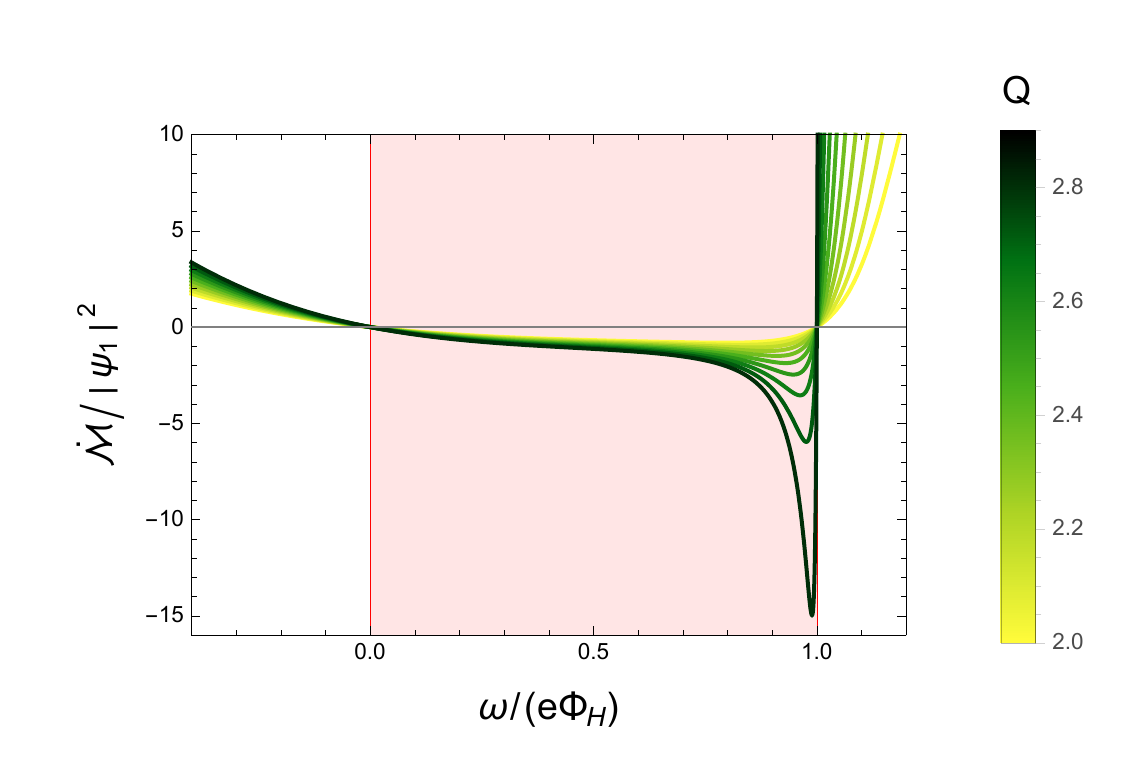}
  \caption{The $\omega/(e\Phi_H)$ dependence of the energy variation $\dot{\mathcal{M}}/|\psi_1|^2$ for the black hole charges $Q=2.0,\,2.1,\,\cdots,2.9$. The superradiant condition is satisfied in the red region.}
    \label{fig:dM_RN}
\end{figure}

\begin{figure}[t]
    \centering
    \includegraphics[width=0.8\linewidth]{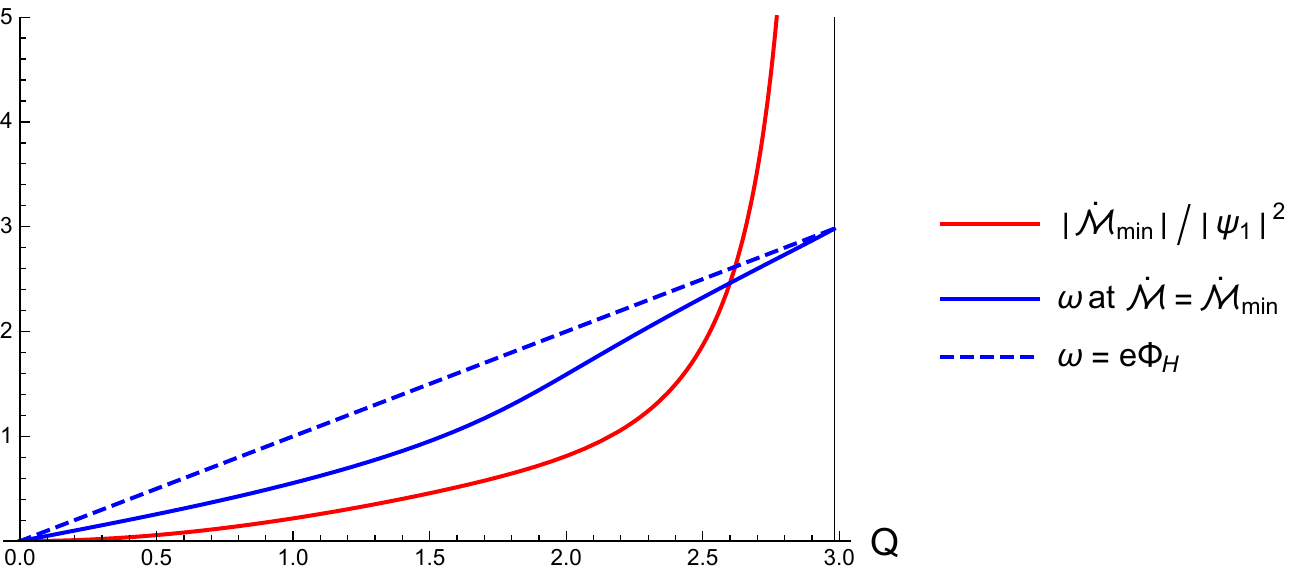}
    \caption{Maximum energy extraction rate with respect to $Q$. The red line is the maximum extracted energy per unit time $|\dot{\mathcal{M}}_{\text{min}}|/|\psi_1|^2$, the blue line is the corresponding frequency, and the blue dashed line shows $\omega= e\Phi_H$. The black grid line at $Q=2.981$ is the location of the onset of instability.}
    \label{fig:max_extraction}
\end{figure}

The peak for each $Q$ in Fig.~\ref{fig:dM_RN} shows that we can extract the maximum energy per unit time $|\dot{\mathcal{M}}_{\text{min}}|$ from the black hole by applying an appropriate wave frequency. The $Q$-dependence of the maximum energy extraction rate is shown in Fig.~\ref{fig:max_extraction}. The red line is the maximum extraction rate $|\dot{\mathcal{M}}_{\text{min}}|/|\psi_1|^2$, the blue line is the corresponding frequency, and the blue dashed line is $\omega= e\Phi_H$. A black vertical line at $Q=2.981$ shows the location of the onset of instability. The behavior of $|\dot{\mathcal{M}}_{\text{min}}|/|\psi_1|^2$ implies that the maximum value of the extractable energy grows rapidly as we get close to the onset of instability. Note that the quasinormal mode that becomes unstable (see appendix~\ref{sec:qnm}) has a respectable residue and affects the spectrum of the response \cite{Kaminski:2009dh}. Also, we see that the negative peak gets closer to the upper bound of the superradiant condition $\omega=e\Phi_H$ as we raise $Q$. 
If we tune the wave frequency so as to follow the blue line in Fig.~\ref{fig:max_extraction} toward the left-bottom, we can extract the mass and the charge of the black hole in the most rapid way. Note that although we can extract the largest energy per unit time by following the blue line, the total amount of the extracted energy is not always the maximum, since it is not an isentropic process (see also Fig.~\ref{fig:MQ}).

\section{Rotational superradiance in AdS}
\label{sec:kerr}

In this section, we consider a neutral scalar field in the Kerr-AdS$_4$ black hole background for an alternative setup of the energy extraction from AdS black holes.

\subsection{Kerr-AdS$_4$ spacetime}

The metric of the Kerr-AdS$_4$ spacetime is given by \cite{Carter:1968ks,Hawking:1998kw,Gibbons:2004uw}
\begin{multline}
 ds^2=-\frac{\Delta}{\rho^2}\left(dt-\frac{a}{\Xi}\sin^2\theta d\phi\right)^2+\frac{\rho^2}{\Delta}dr^2\\
+\frac{\rho^2}{\Delta_\theta}d\theta^2
+\frac{\Delta_\theta\sin^2\theta}{\rho^2}\left(adt-\frac{r^2+a^2}{\Xi}d\phi\right)^2\ ,
\label{KAdS}
\end{multline}
where
\begin{equation}
\begin{split}
&\Delta=(r^2+a^2)(1+r^2)-2Mr\ ,\quad
\Delta_\theta=1-a^2\cos^2\theta\ ,\\
&\rho^2=r^2+a^2\cos^2\theta\ ,\quad
\Xi=1-a^2\ ,
\end{split}
\end{equation}
and we set the AdS radius unity: $L=1$.
The event horizon is located at $r=r_+$ given by the largest root of $\Delta(r)=0$. 
Note that the above metric is written in the rotating frame at infinity.

The metric can be rewritten in the non-rotating frame at infinity.
We introduce new coordinates $(t',\phi',\theta',\phi')$ defined as~\cite{Hawking:1998kw}
\begin{equation}
 t'=t\ ,\quad \phi'=\phi+at\ ,\quad r'\cos\theta' = r \cos\theta\ ,\quad \Xi r'^2 = \Delta_\theta r^2 + a^2\sin^2\theta\ .
\label{nonrot}
\end{equation}
However, directly transforming the metric \eqref{KAdS} is cumbersome, and hence we focus only on the asymptotic form near the asymptotic infinity $r'\to \infty$. There, the original coordinates $(r,\theta)$ can be expressed by $(r',\theta')$ as 
\begin{equation}
r=\frac{1}{h(\theta')} \,r' + \mathcal{O}(1)\ ,\quad 
\cos \theta =  h(\theta')\cos\theta' + \mathcal{O}(1/r'^2)\ ,
\label{nonrot_inf}
\end{equation}
where
\begin{equation}
 h(\theta')^{-2}\equiv 1- a^2 \sin^2\theta'\ .
\end{equation}
It is easy to check that the following formula holds:
\begin{equation}
 d(h(\theta')\cos\theta')=\Xi \,h(\theta')^3 d(\cos\theta')\ .
\label{dformula}
\end{equation}
This will be convenient when we change integration variables later.
In terms of the new coordinate system, the asymptotic form of the metric becomes
\begin{equation}
 ds^2\simeq -(1+r'{}^2)dt'^2 + \frac{dr'^2}{1+r'{}^2} + r'{}^2(d\theta'{}^2+\sin^2\theta' d\phi'{}^2)\ .
\end{equation}

The mass, angular momentum and entropy of the Kerr-AdS$_4$ are given by~\cite{Gibbons:2004ai}
\begin{equation}
 \mathcal{M}=\frac{8\pi M}{\kappa^2 \Xi^2}\ ,\quad \mathcal{J}=\frac{8\pi Ma}{\kappa^2\Xi^2}\ ,\quad
\mathcal{S}=\frac{8\pi^2(r_+^2+a^2)}{\kappa^2\Xi}\ .
\label{KerrMJS}
\end{equation}
The Hawking temperature and angular velocity of the horizon are 
\begin{equation}
 T=\frac{r_+(1+a^2+3r_+^2-a^2/r_+^2)}{4\pi(r_+^2+a^2)}\ ,\quad \Omega = \frac{a(1+r_+^2)}{r_+^2+a^2}\ .
\end{equation}
They satisfy the first law of black hole mechanics:
\begin{equation}
 d\mathcal{M} = Td\mathcal{S} + \Omega d\mathcal{J}\ .
\end{equation}

\subsection{Scalar field perturbation of Kerr-AdS$_4$}

Let us consider the massive Klein-Gordon equation $\Box \Psi=\mu^2 \Psi$ in the Kerr-AdS$_4$.
We decompose the scalar field as 
\begin{equation}
 \Psi=e^{-i\omega t' + im\phi'}R_{\omega l m}(r)S_{\omega l m}(\cos\theta)=e^{-i(\omega-am) t + im\phi}R_{\omega l m}\,(r)S_{\omega l m}(\cos\theta)
\end{equation}
where we define the frequency $\omega$ and the azimuthal mode number $m$ in terms of the coordinates in  the non-rotating frame.
The Klein-Gordon equation is separated into
\begin{equation}
 \bigg[\frac{d}{dr} \Delta \frac{d}{dr} +\frac{(r^2+a^2)^2}{\Delta}\left(\omega -\frac{a(r^2+1)}{r^2+a^2}m\right)^2
-\mu^2 r^2 \bigg]R_{\omega l m}(r)=A_{\omega l m}R_{\omega l m}(r)\ ,
\label{Req}
\end{equation}
and
\begin{multline}
 \bigg[\frac{1}{\sin\theta} \frac{d}{d\theta} \Delta_\theta \sin\theta \frac{d}{d\theta}
-\frac{a^2\sin^2\theta}{\Delta_\theta}\left(\omega-\frac{\Delta_\theta}{a\sin^2\theta}m\right)^2\\
-\mu^2 a^2\cos^2\theta\bigg] S_{\omega l m}(\cos \theta) = -A_{\omega l m}S_{\omega l m}(\cos \theta)\ ,
\label{Seq}
\end{multline}
where $A_{\omega l m}$ is the separation constant and determined by solving the eigenvalue problem given by Eq.(\ref{Seq}). 
The eigenvalue is parametrized by the integers $l$ and $m$ satisfying $l\geq |m|$ as well as $\omega$. (The eigenfunction becomes $S_{\omega l m}(\theta)e^{im\phi}\to Y_{lm}(\theta,\phi)$ in $a\to 0$, where $Y_{lm}(\theta,\phi)$ are the spherical harmonics.)
Since the differential operator in the left hand side of Eq.(\ref{Seq}) is hermitian under the inner product $(f,g)\equiv \int^1_{-1} d(\cos\theta) f^\ast(\theta)g(\theta)$, the eigenfunctions $S_{\omega l m}$ are orthogonal.
We normalize them such that
\begin{equation}
 (S_{\omega l m},S_{\omega l' m})=\frac{\Xi}{2\pi}\delta_{ll'}\ .
\label{SSorth}
\end{equation}
In the following, we will fix $(\omega,l,m)$ and omit tedious subscripts as
\begin{equation}
 R_{\omega l m}(r) = R(r)\ ,\quad 
 S_{\omega l m}(\cos \theta)=S(\cos \theta)\ ,\quad
 A_{\omega lm}=A\ .
\end{equation}

Once $A$ is determined, we turn to the radial part of the Klein-Gordon equation \eqref{Req}. With a rescaled variable $\psi(r)=\sqrt{r^2+a^2}R(r)$, it is rewritten as
\begin{equation}
 \left[-\frac{d^2}{dr_\ast^2} + V(r)\right]\psi  = \left(\omega-\frac{a(r^2+1)}{r^2+a^2}m\right)^2 \psi\ .
\end{equation}
where $r_\ast = \int_\infty^r dr' (r'{}^2+a^2)/\Delta(r')$ and 
\begin{multline}
 V=\frac{\Delta}{(r^2+a^2)^4}\{2 r^6 + 5 a^2 r^4 + 2 M r^3 + a^2 (3 a^2+ 1) r^2\\
- 4 M a^2 r  +a^4 + (r^2+a^2)^2 (A + r^2  \mu^2)\}\ .
\end{multline}
Hereafter, we consider the case that $\mu^2=-2$ for simplicity. Then, near the AdS boundary, the asymptotic form of the solution is of the form
\begin{equation}
 \psi =\psi_1  + \frac{\psi_2}{r} + \cdots \ .
\label{psiexpandkerr}
\end{equation}
Near the horizon, we impose the ingoing wave condition as
\begin{equation}
 \psi(z)\simeq C e^{-i(\omega-m\Omega) r_\ast}\ ,
\label{ingoingKerr}
\end{equation}
where $C$ is a complex constant. The conservation of the Wronskian~(\ref{Wronsdef}) implies that
\begin{equation}
 \psi_1\psi_2^\ast-\psi_1^\ast\psi_2=-2i(\omega-m\Omega)|C|^2\ .
\label{WronsKerr}
\end{equation}

The coefficients $\psi_1$ and $\psi_2$ in Eq.(\ref{psiexpandkerr}) are related to the source and response in the boundary theory, but we need to be careful about the nontrivial coordinate transformation~(\ref{nonrot}). 
In the coordinates in the non-rotating frame $(t',r',\theta',\phi')$, the expansion of the scalar field near the asymptotic infinity becomes 
\begin{equation}
\begin{split}
 \Psi&=e^{-i\omega t' +im\phi'}\frac{S(\cos\theta)}{\sqrt{r^2+a^2}}\left(\psi_1  + \frac{\psi_2}{r} + \cdots \right)\\
&=e^{-i\omega t' +im\phi'}S(h(\theta')\cos\theta')h(\theta') \left(\frac{\psi_1}{r'} + \frac{h(\theta')\psi_2}{r'{}^2}+\cdots\right)\ ,
\end{split}
\end{equation}
where Eq.(\ref{nonrot}) was used in the second equality.
Then, the source and response are given by
\begin{equation}
\begin{split}
&J(t',\theta',\phi') = e^{-i\omega t' +im\phi'}S(h(\theta')\cos\theta')h(\theta') \psi_1 \ ,\\
&\langle \mathcal{O}(t',\theta',\phi')\rangle =e^{-i\omega t' +im\phi'}S(h(\theta')\cos\theta')h(\theta')^2 \psi_2 \ . 
\end{split}
\label{JOKerr}
\end{equation}

\subsection{Energy extraction through superradiance}

In the Kerr-AdS$_4$ background, we use the Ward-Takahashi identity~(\ref{dT}) with a vanishing Maxwell field, $A_{i}=0$. 
Calculations are parallel to the case of the RNAdS$_4$.
The change of the energy per unit time is given by
\begin{equation}
\dot{\mathcal{M}}= -\int d\Omega (\langle \mathcal{O}\rangle^\ast \partial_{t'}J + \langle \mathcal{O}\rangle \partial_{t'}J^\ast) \ ,
\label{dotM1}
\end{equation}
where $\int d\Omega = \int_0^{2\pi} d\phi' \int^1_{-1} d(\cos\theta')$.
Note that the energy is defined in the non-rotating frame at infinity.
Meanwhile, the angular momentum is 
defined on a spacelike hypersurface $\Sigma$ by $-\int d\Sigma T_{ij}n^i \xi^j$ with $\xi=\partial_{\phi'}$. 
In a similar way as Eq.(\ref{Mchange}), we obtain
the change of the angular momentum per unit time as
\begin{equation}
 \dot{\mathcal{J}}=\int d\Omega (\langle \mathcal{O}\rangle^\ast \partial_{\phi'} J + \langle \mathcal{O}\rangle \partial_{\phi'} J^\ast) \ .
\end{equation}

These expressions can be translated to the superradiant condition by making use of the Wronskian.
Using Eq.(\ref{JOKerr}) to (\ref{dotM1}), we obtain
\begin{equation}
\begin{split}
\dot{\mathcal{M}}&= i\omega \int_0^{2\pi} d\phi' \int^1_{-1} d(\cos\theta') h(\theta')^3 |S(h(\theta')\cos\theta')|^2 (\psi_1 \psi_2^\ast - \psi_1^\ast \psi_2) \\
&= i\omega  (\psi_1 \psi_2^\ast - \psi_1^\ast \psi_2) = 2 \omega (\omega - m\Omega)|C|^2\ ,
\end{split}
\end{equation}
where we used Eqs.(\ref{dformula}) and $(\ref{SSorth})$ in the second equality and the conservation of the Wronskian~(\ref{WronsKerr}) in the last equality. 
Similarly, we have
\begin{equation}
 \dot{\mathcal{J}}=i m  (\psi_1 \psi_2^\ast - \psi_1^\ast \psi_2) = 2 \omega (\omega - m\Omega)|C|^2\ .
\end{equation}
We find that $\dot{\mathcal{M}}$ becomes negative for
\begin{equation}
 0<\omega<m \Omega \quad (m\Omega>0)\ ,\qquad m\Omega < \omega <0\quad(m\Omega<0)\ .
\label{SRcondKerr}
\end{equation}

Thus, we can extract energy from the thermal state dual to the Kerr-AdS$_4$ by applying a source with the frequency satisfying the above condition.
In asymptotically flat spacetime, the above inequality appears as the ordinary superradiant condition.
In the asymptotically AdS spacetime, it can be regarded as the condition for the energy extraction from the rotating black hole.

\begin{figure}[t]
\centering
\includegraphics[scale=0.72]{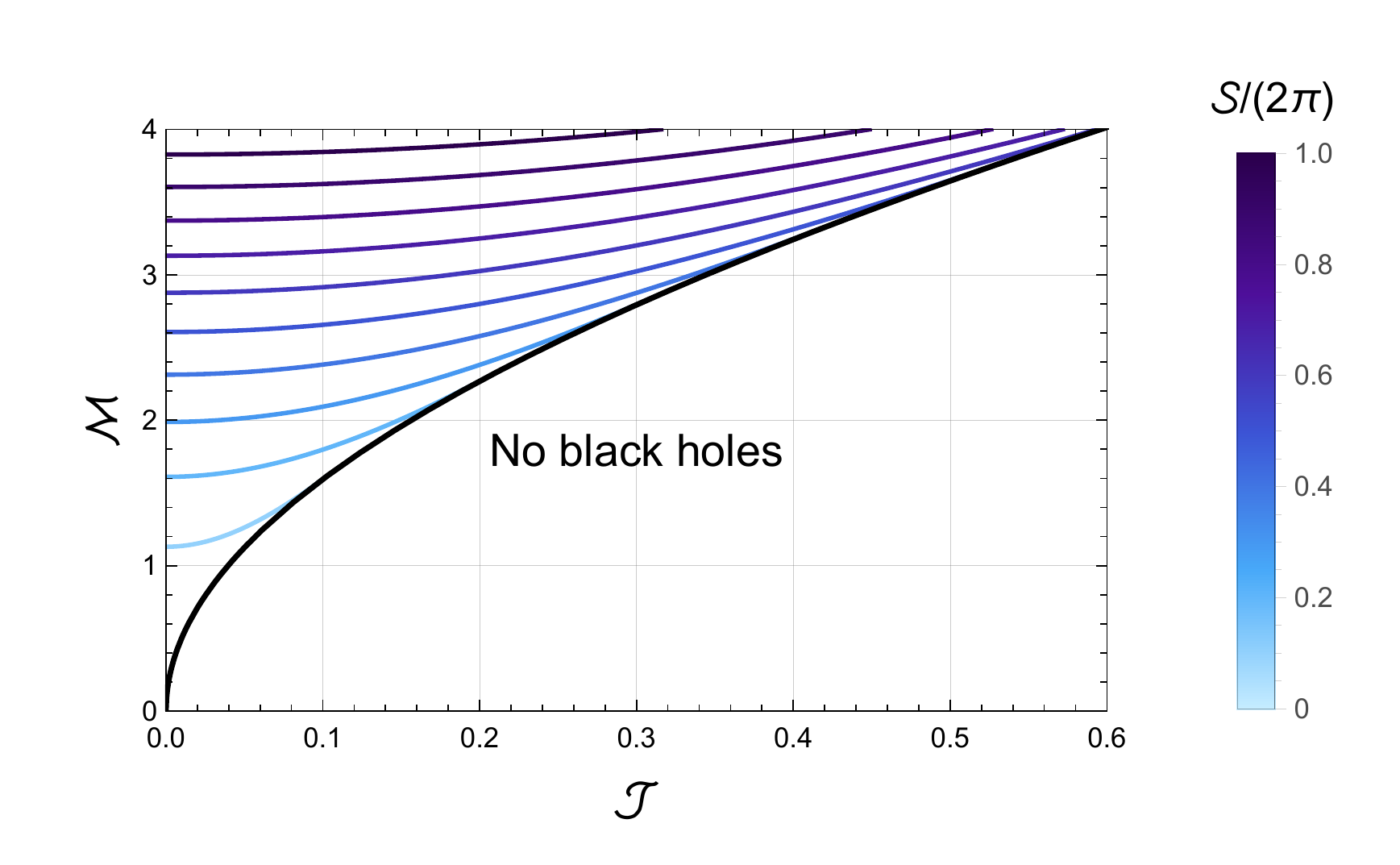}
\caption{Contours of the entropy of Kerr-AdS$_4$}
\label{fig:MJ_constS}
\end{figure}

The first law of black hole mechanics implies the change of the entropy to be
\begin{equation}
 T\dot{\mathcal{S}}=\dot{\mathcal{M}}- \Omega \dot{\mathcal{J}}=2(\omega-m\Omega)^2 |C|^2\ .
\label{dotSKerr}
\end{equation}
This is always positive and consistent with Hawking's area theorem. For the same reason as argued in section~\ref{energyextract}, the most efficient energy extraction can be realized by an isentropic process $\dot{\mathcal{S}}=0$ as in Fig.~\ref{fig:MJ_constS}, where eliminating $r_+$ and $a$ from \eqref{KerrMJS} gives $\mathcal{M}$ as a function of $(\mathcal{S},\mathcal{J})$ as
\begin{equation}
\mathcal{M}(\mathcal{S},\mathcal{J})=\frac{\sqrt{8\pi^2+\mathcal{S}}}{4\pi^2\sqrt{2\mathcal{S}}}\sqrt{32\pi^4\mathcal{J}^2+8\pi^2\mathcal{S}+\mathcal{S}^3} \ ,
\end{equation}
where we set $\kappa^2=1$ for simplicity.
Let the initial mass and angular momentum  be $\mathcal{M}_\textrm{ini}$ and $\mathcal{J}_\textrm{ini}$. Then, the initial entropy $\mathcal{S}_\textrm{ini}$ is obtained by solving $\mathcal{M}_\textrm{ini}=\mathcal{M}(\mathcal{S}_\textrm{ini},\mathcal{J}_\textrm{ini})$. For the isentropic process, the final mass is given by $\mathcal{M}_\textrm{fin}=\mathcal{M}(\mathcal{S}_\textrm{ini},0)$. The maximum energy that can be extracted is $\Delta \mathcal{M}=\mathcal{M}_\textrm{ini}-\mathcal{M}(\mathcal{S}_\textrm{ini},0)$.

\subsection{Detailed calculation of the energy extraction}

In this subsection, we numerically calculate the variation of the mass $\dot{\mathcal{M}}$ for different wave frequencies $\omega$ and black holes spins $a$. We will again show that the condition \eqref{dotM} is satisfied indeed. We take units in which the AdS radius is unity and normalize the variation of the mass by the amplitude of the source: $\dot{\mathcal{M}}/|\psi_1|^2$. 
In the following, the azimuthal mode number is fixed to $m=1$, and the separation constant $A_{\omega lm}$ is taken as the lowest eigenvalue of Eq.~\eqref{Seq}; that is, $m=1,\,l=1$. Other choices of $m,\,l$ also give qualitatively the same behavior.

According to \cite{Uchikata:2009zz}, the Kerr-AdS$_4$ black holes are unstable against the scalar perturbation for a small horizon radius. In particular, the onset of the instability appears for $r_+\lesssim 0.1$ in the case of $l=1,m=1$. Instability of the small Kerr-AdS$_4$ black holes has also been also studied in \cite{Cardoso:2004hs,Cardoso:2006wa}. Since we will consider a larger black hole $r_+=1$ in order to avoid numerical difficulty in the following, we will not see the divergence behavior due to the small Kerr-AdS$_4$ instability unlike the calculations in the RNAdS$_4$ case.

\begin{figure}[t]
    \centering
    \includegraphics[width=0.8\linewidth]{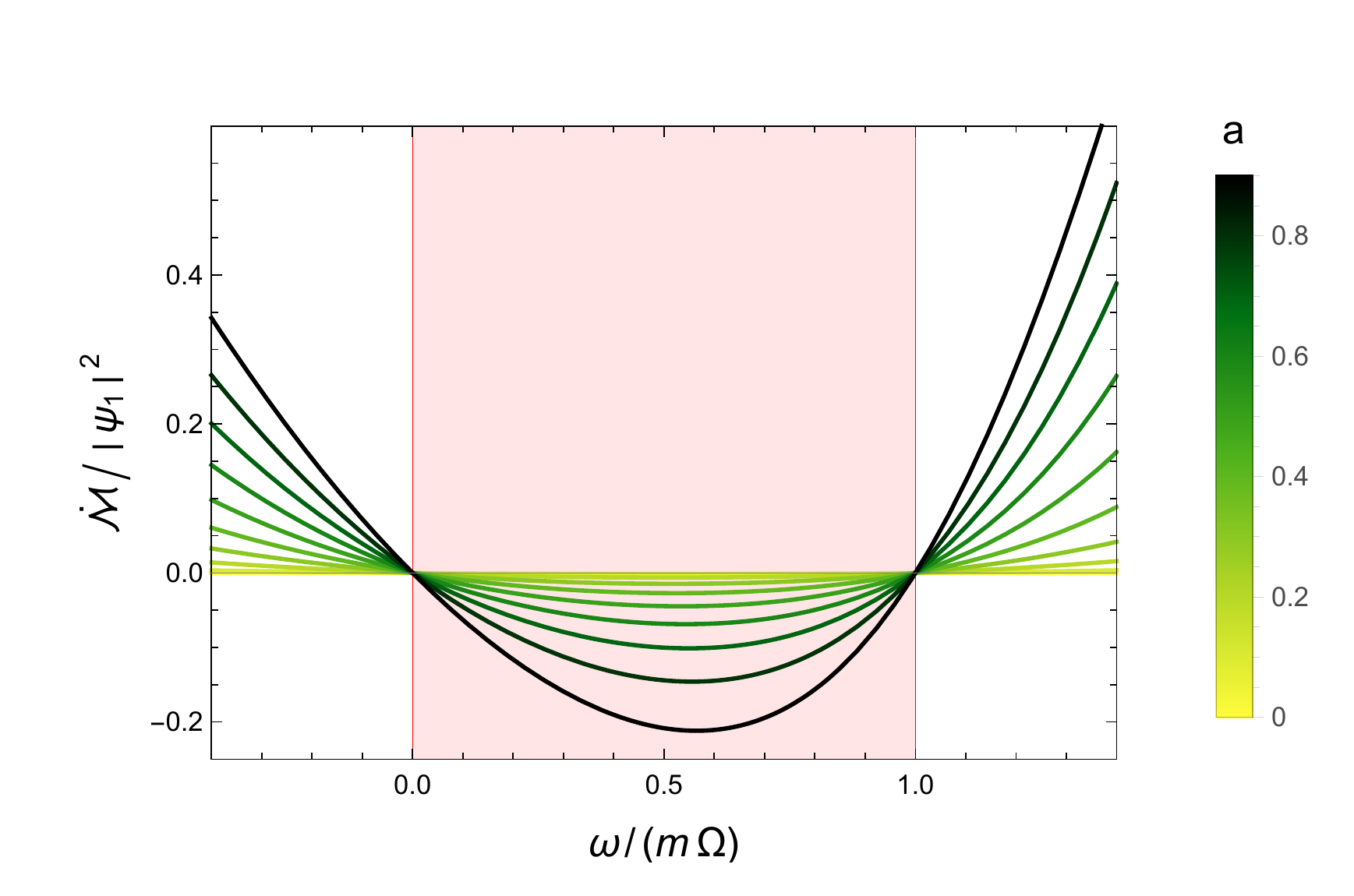}
  \caption{The $\omega/(e\Phi_H)$ dependence of the energy variation $\dot{\mathcal{M}}/|\psi_1|^2$ for the black hole spins $a=0,\,0.1,\, \cdots, 0.9$. The superradiant condition is satisfied in the red region.}
    \label{fig:dM_Kerr}
\end{figure}

In Fig.~\ref{fig:dM_Kerr}, we show the $\omega$-dependence of the rescaled variation of the mass $\dot{\mathcal{M}}/|\psi_1|^2$ with respect to $a=0.,\,0.1,\, \cdots, 0.9$. 
We also use the horizontal axis rescaled as $\omega/(m\Omega)$, where the superradiant condition is satisfied for $0<\omega/(m\Omega)<1$ as shown by the red region. We find that $\dot{\mathcal{M}}$ is negative in the frequency band. That means that we can extract the energy from the rotating black hole when the superradiant condition is satisfied. The maximal value of the energy extraction increases for a larger spin and reaches the upper limit at $a\lesssim 1$. As described earlier, there is no divergent behavior since the onset of the instability does not appear in the parameters we are considering.

\section{Conclusion}
\label{sec:sum}

In this paper, we studied the superradiance for the charged scalar field perturbation of RNAdS$_4$ and the neutral scalar field perturbation of Kerr-AdS$_4$ when we apply sources with monochromatic frequencies from the AdS boundary. We discussed the energy extraction due to the superradiance. We derived the expressions of the change rates for mass, electric charge, and angular momentum. The mass is shown to be extracted by the perturbation. That means that the work done by the source at the AdS boundary is negative. Numerical results have been obtained showing the energy extraction for frequencies in the range satisfying the superradiant condition. 

The phenomena of the superradiance found in the AdS bulk can be mapped to that in the dual field theory. 
The RNAdS$_4$ and Kerr-AdS$_4$ correspond to equilibrium thermal states having finite charge and angular momentum, respectively. 
Applying external fields with frequencies satisfying Eq.(\ref{SRcond}) or Eq.(\ref{SRcondKerr}), we can extract energy even though the states are initially in equilibrium. In particular, if we fine-tune the frequency, the process becomes isentropic. 
Since the isentropic process is reversible, we can also add the energy keeping the entropy. 
Through this process, the energy is stored in the black holes. In other words, black holes play the role of a ``battery''.

Even for the charged scalar field in the RNAdS$_4$ with the flat horizon, we find the energy extraction associated with the superradiant condition. 
The instability for a large black hole is generically known as the instability due to the near horizon AdS$_2$ region of the RNAdS$_4$ and hence called the AdS near horizon instability. 
Focusing on the ability of the energy extraction, however, we have observed the superradiance even for large black holes.

In this work, we considered applying sources with monochromatic frequencies to extract black hole energy. It would be also interesting to study if the superradiance can be identified when a scalar field quench with a general profile is applied as a source. Another future direction will be to consider the application to the Kerr-Newman-AdS$_4$ black holes. Taking into account the backreaction of the matter fields and considering nonlinear effects to the superradiant energy extraction will be also important.

\begin{acknowledgments}
We would like to thank Shunichiro Kinoshita for useful discussions and comments.
The work of T.~I.~was supported in part by JSPS KAKENHI Grant Number JP19K03871.
Y.~K.~would like to thank the ``Nagoya University 
Interdisciplinary Frontier Fellowship'' supported by JST and Nagoya University.
This work of K.~M.~was supported in part by JSPS KAKENHI Grant Nos. 20K03976, 21H05186 and 22H01217.
\end{acknowledgments}

\appendix

\section{Ward-Takahashi identity}
\label{sec:WTI}

In this appendix, we summarize the derivation of the Ward-Takahashi identities~(\ref{dj}) and (\ref{dT}) (see e.g.~Ref.~\cite{Balasubramanian:1999re,Ashtekar:1999jx,deHaro:2000vlm,Skenderis:2002wp}).

The full bulk action of the Einstein-Maxwell-charged scalar system in asymptotically AdS$_4$ spacetime is written as
\begin{multline}
 S = \frac{1}{2\kappa^2}\int d^4x \sqrt{-g} \left[R-6 - \frac{1}{4}F_{\mu\nu}F^{\mu\nu}\right] 
+ \int d^4x \sqrt{-g} (-|D\Psi|^2 + 2|\Psi|^2)\\
+ \frac{1}{\kappa^2}\int_{z=\epsilon} d^3x \sqrt{-h} \left[K-2-\frac{1}{2}R^{(h)}\right]
- \int_{z=\epsilon} d^3x \sqrt{-h} |\Psi|^2\ ,
\label{fullaction}
\end{multline}
where $h_{ij}$ is the induced metric on the cutoff surface at $z=\epsilon$, and $h=\textrm{det}(h_{ij})$. $R^{(h)}$ is the three-dimensional Ricci scalar with respect to $h_{ij}$.
The extrinsic curvature is defined as $K_{ij}=(\nabla_\mu l_\nu) e^\mu_i e^\nu_j$, where $l_\mu$ and $e^\mu_i$ are the unit normal vector and projection tensor of the hypersurface of $z=\epsilon$.
The third bracket in \eqref{fullaction} contains the boundary terms for the gravitational action, including the Gibbons-Hawing term and counterterms~\cite{Brown:1992br,Balasubramanian:1999re}. 
The last term is the counterterm for the charged scalar field, renormalizing the on-shell action in the limit $\epsilon\to 0$~\cite{Hartnoll:2008kx}. 

Because the induced metric $h_{ij}$ diverges for $\epsilon\to 0$, the rescaled induced metric regular at the AdS boundary is introduced as
\begin{equation}
 \gamma_{ij} = \lim_{\epsilon\to 0} z^2 h_{ij}|_{z=\epsilon}\ .
\end{equation}

In the main text of the paper, we consider the limit of a probe scalar field. The scalar field is decoupled from the gravity and Maxwell field in the limit $\kappa^2\to 0$ while keeping the gravity and Maxwell fields in the same order. In fact, the action \eqref{fullaction} is written in such a normalization for the Maxwell field, and the probe approximation is justified.

The expectation value of the operator dual to the scalar field is common to both the probe scalar limit and the fully back-reacted system. The scalar field is expanded near infinity as in Eq.(\ref{phiexpand}).
Substituting this expansion into the variation of the on-shell action with respect to $\Psi(t,x,y)$, we obtain 
\begin{equation}
 \delta_\Psi S = \int dt dx dy [ \Psi_2 \delta \Psi_1^\ast + \Psi_2^\ast \delta \Psi_1 ] \ .
\end{equation}
The counterterm of the scalar field in Eq.(\ref{fullaction}) cancels divergent terms in the bare action, and the variational principle is well-defined when $\Psi_1 (t,x,y)$ is varied at the AdS boundary.\footnote{We employ the standard quantization where the $\Psi_1$ is varied. For the mass squared $\mu^2=-2$, the alternative quantization that uses the variation for $\Psi_2$ is also possible, but we do not consider the latter in this paper.}
From the dictionary of the AdS/CFT correspondence \cite{Gubser:1998bc,Witten:1998qj}, $J\equiv \Psi_1$ is regarded as the source of the scalar operator $\mathcal{O}$ dual to $\Psi$ in the boundary theory. 
The response with respect to the source $J$ is given by
\begin{equation}
 \langle \mathcal{O}\rangle = \frac{1}{\sqrt{-\gamma}}\frac{\delta S}{\delta J^\ast} = \Psi_2\ .
 \label{vevO}
\end{equation}

We regard the bulk on-shell action \eqref{fullaction} as the functional of $\gamma_{ij}$, $A_i$, and  $J=\Psi_1$,
where $A_i$ is the induced Maxwell field on $z=\epsilon$.
Along with \eqref{vevO}, the boundary stress tensor $\langle T_{ij} \rangle$ and the electric current $\langle j^i \rangle$ are defined by 
\begin{equation}
\langle T_{ij} \rangle = -\frac{2}{\sqrt{-\gamma}}\frac{\delta S}{\delta \gamma^{ij}}\ ,\quad
\langle j^i    \rangle = \frac{1}{\sqrt{-\gamma}}\frac{\delta S}{\delta A_i}\ .
\end{equation}
The general variation of the on-shell action is then written as
\begin{equation}
 \delta S = \int d^3x \sqrt{-\gamma}\left[-\frac{1}{2}\langle T_{ij} \rangle \delta \gamma^{ij} + \langle j^i \rangle  \delta A_i + \langle \mathcal{O}\rangle \delta J^\ast +\langle \mathcal{O}\rangle^\ast \delta J\right]\ .
 \label{deltaS}
\end{equation}

The $U(1)$-gauge invariance of the onshell action $\delta_{U(1)} S = 0$ implies the divergence formula for the electric current~(\ref{dj}). 
Let us take the variation $\delta$ as the $U(1)$-gauge transformation as
\begin{equation}
 \delta_{U(1)}\gamma^{ij}=0\ ,\quad \delta_{U(1)} A_i = \partial_i \lambda\ ,\quad \delta_{U(1)} J=ie \lambda J\ ,
\end{equation}
where $\lambda$ is the arbitrary function. Then, we obtain(\ref{dj}) from (\ref{deltaS}).

Similarly, we consider the coordinate transformation $\delta_\textrm{diff}$ given by
\begin{equation}
 \delta_\textrm{diff} \gamma^{ij} = -\nabla^i \xi^j - \nabla^j \xi^i\ ,\quad \delta_\textrm{diff} A_i = \xi^j \nabla_j A_i+A_j \nabla_i \xi^j\ ,\quad
 \delta_\textrm{diff} J=\xi^j \partial_j J\ ,
\end{equation}
where $\xi^i$ is the arbitrary vector field.
From the diffeomorphism invariance of the on-shell action $\delta_\textrm{diff} S = 0$, we obtain
\begin{equation}
 \nabla^i \langle T_{ij} \rangle 
=\langle j^i \rangle  \nabla_j A_i-\nabla_i(\langle j^i \rangle A_j) + \langle \mathcal{O}\rangle \partial_j J^\ast +\langle \mathcal{O}\rangle^\ast \partial_j J \ .
\end{equation}
Rewriting $\nabla_i \langle j^i \rangle$ by using Eq.(\ref{dj}), we obtain the divergence formula for the energy momentum tensor~(\ref{dT}).

\section{Backreaction of the superradiance}
\label{sec:BReac}

We can compute the leading contribution to the change of mass and charge of the black hole by a second-order perturbation. (See also \cite{Brito:2015oca} for the second-order perturbation of asymptotically flat charged black holes.) The probe approximation of the scalar field is valid in the limit of $\kappa^2 \to 0$, and the scalar field decouples from the gravity and Maxwell field. 
We then take into account the backreaction of the scalar field to these fields perturbatively. 

The scalar field is quadratic and gives an $\mathcal{O}(\kappa^2)$ contribution to the equations for the gravity and Maxwell field.
The Einstein equations obtained from the full action~(\ref{fullaction}) are 
\begin{equation}
 R_{\mu\nu}-\frac{1}{2}g_{\mu\nu}R -3 g_{\mu\nu}= \frac{1}{2}T_{\mu\nu}^\textrm{Maxwell} +  \kappa^2\, T_{\mu\nu}^\textrm{Scalar}\ ,
\label{Estnfull}
\end{equation}
where
\begin{align}
 &T_{\mu\nu}^\textrm{Maxwell}=F_{\mu\rho}F_{\nu}{}^\rho-\frac{1}{4}g_{\mu\nu}F_{\rho\sigma}F^{\rho\sigma}\ ,\\
 &T_{\mu\nu}^\textrm{Scalar}=2D_{(\mu} \Psi D_{\nu)}\Psi^\ast + g_{\mu\nu}(-|D\Psi|^2+2|\Psi|^2)\ .
\end{align}
The Maxwell equations are 
\begin{equation}
 \nabla^\nu F_{\mu\nu} = 2 \kappa^2 \, j_\mu^\textrm{Scalar}\ ,\quad j_\mu^\textrm{Scalar} = ie(\Psi D_\mu\Psi^\ast - \Psi^\ast D_\mu\Psi)\ .
\label{MXfull}
\end{equation}

We consider the perturbation of the metric and Maxwell field as
\begin{equation}
 g_{\mu\nu}\to g_{\mu\nu} + \kappa^2 \delta g_{\mu\nu}\ ,\quad A_\mu\to A_\mu + \kappa^2 \delta A_\mu\ .
\end{equation}
We will focus on the $\mathcal{O}(\kappa^2)$ contribution of the scalar field to these fields.
In $j_\mu^\textrm{Scalar}$ and $T_{\mu\nu}^\textrm{Scalar}$, we can take $\Psi$ as the solution of the Klein-Gordon equation~(\ref{KG}) in the fixed background metric $g_{\mu\nu}$ and Maxwell field $A_\mu$.

For simplicity, we assume that the scalar field is homogeneous in the $(x,y)$-space, i.e., $\Psi=\Psi(t,z)$.
Near the AdS boundary, we can expand the scalar field as 
\begin{equation}
 \Psi(t,z)=\sum_{n=1}^\infty \Psi_n(t) z^n\ .
\end{equation}
For $n\geq 3$, $\Psi_{n}(t)$ is determined by the equation of motion~(\ref{KG}) as
\begin{equation}
 \Psi_3(t)= \frac{\ddot{\Psi}_1}{2}\ ,\quad 
 \Psi_4(t)= \frac{ieQ}{3}\ddot{\Psi}_1 + \frac{M}{3}\Psi_1 + \frac{1}{6}\ddot{\Psi}_2\ ,
\end{equation}
and so on. Because of the homogeneity of the scalar field, we can assume that $\delta g_{\mu\nu}$ and $\delta A_\mu$ are also homogeneous:
$\delta g_{t a} = \delta g_{z a} = \delta g_{xy} =\delta A_a = 0$ and $\delta g_{xx}=\delta g_{yy}$ where $a=x,y$. 

Note that there are gauge freedom in the perturbation: $\delta g_{\mu\nu} \to \delta g_{\mu\nu}-\nabla_\mu \xi_\nu-\nabla_\nu \xi_\mu$ and $\delta A \to \delta A + d\lambda$ where $\xi = \xi_t(t,z)dt + \xi_z(t,z)dz$ and $\lambda=\lambda(t,z)$. Using these, we can further impose $\delta g_{tz}=g_{xx}=0$ and $\delta A_z=0$. Then, the perturbation variables can be written as
\begin{equation}
 \delta g_{\mu\nu}dx^\mu dx^\nu = z \left(\alpha(t,z) dt^2 + \frac{\beta(t,z)}{F^2}dz^2\right)\ ,\quad \delta A = -z \gamma(t,z) dt \ .
\end{equation}

In $\mathcal{O}(\kappa^2)$, we obtain a set of coupled equations for $\alpha(t,z), \beta(t,z)$, and $\gamma(t,z)$. Taking the $\mathcal{O}(\kappa^2)$ part of the Maxwell equations~(\ref{MXfull}), we find
\begin{align}
&zF\left\{z  A_t' \left(\frac{z^3(\beta-\alpha)}{2F}\right)'  + (z^2\gamma')'\right\} + z^5 A_t'' \beta = 2J_t^\textrm{Scalar}\ ,\label{M1}\\
&\frac{z^2}{F}\left\{(z\dot{\gamma})'  + \frac{z^3 A_t' }{2F}(\dot{\beta}-\dot{\alpha})\right\}=2J_z^\textrm{Scalar}\ .
\end{align}
From the Einstein equations~(\ref{Estnfull}), we obtain
\begin{align}
&-z^2F \bigg\{ \beta'
- \frac{z^3 A_t'{}^2}{4F} \beta 
+\frac{zF' -3(F-1)}{zF}\alpha- \frac{A_t'}{2}(z\gamma)'\bigg\} =T_{tt}^\textrm{Scalar}\ ,\\
&- \frac{z^2}{F}\dot{\beta}= T_{tz}^\textrm{Scalar}\ ,\label{dotbetaeq}\\
&
\frac{z^2}{F} \alpha'
+\frac{1}{4F^2}\{z^5 A_t'^2 
+ 12 z F -4 z^2 F' \}\alpha
- \frac{ 3 z}{F^2} \beta  - \frac{z^2 A_t'}{2F} (z \gamma)' =T_{zz}^\textrm{Scalar}\ ,\\
&
- \frac{z^3}{2F^2} \ddot{\beta}
- \frac{z^3}{2} \alpha'' 
- 2 z^2 \alpha'+z^2 \beta'  - \frac{z^3F'}{4 F}(\beta-\alpha)' \nonumber \\
&\hspace{3cm}
+ \frac{z^2}{4} (\frac{ z F'{}^2}{F^2} 
- \frac{2 z F''}{F}   
+ \frac{F'}{F}) (\beta-\alpha) 
+\frac{z^5 A_t'^2}{4F} (\beta-\alpha)\nonumber\\
&\hspace{6cm}+ \frac{z^2A_t'}{2} (z\gamma)' 
 = T_{xx}^\textrm{Scalar}= T_{yy}^\textrm{Scalar}\ .\label{E5}
\end{align}

We solve these equations near the AdS boundary. The perturbation variables $(\alpha,\beta,\gamma)$ can be expanded as
\begin{equation}
 \alpha(t,z)=\sum_{n=0}^\infty \alpha_n(t)z^n\ ,\quad 
\beta(t,z)=\sum_{n=-1}^\infty \beta_n(t)z^n\ ,\quad 
\gamma(t,z)=\sum_{n=0}^\infty \gamma_n(t)z^n\ .
\end{equation}
Substituting these expressions into Eqs.(\ref{M1})-(\ref{E5}), we can determine $\alpha_n,\beta_n,\gamma_n$ order by order. For example, we find
\begin{equation}
\begin{split}
 &\beta_{-1} = -|\Psi_1|^2\ ,\quad
 \dot{\alpha}_0 = \frac{1}{3}\{\Psi^\ast_1  \dot{\Psi}_2 + \Psi_1 \dot{\Psi}_2^\ast  - 2 (\Psi^\ast_2  \dot{\Psi}_1 + \Psi_2 \dot{\Psi}_1^\ast )\}\ ,\\
&\beta_0 = -\frac{4}{3}(\Psi_1^\ast \Psi_2 + \Psi_1 \Psi_2^\ast ) + \alpha_0\ ,\quad 
\dot{\gamma}_0 = -2ie (\Psi_1^\ast \Psi_2 - \Psi_1 \Psi_2^\ast )\ ,\\
&\alpha_1= \frac{1}{4}(\ddot{\Psi}_1\Psi_1^\ast  - 2 \dot{\Psi}_1\dot{\Psi}_1^\ast  + \Psi_1\ddot{\Psi}_1^\ast -2 Q \gamma_0 )\ ,\\
&\beta_1=-\frac{1}{2}
(\Psi_1^\ast \ddot{\Psi}_1  + 2 \dot{\Psi}_1^\ast\dot{\Psi}_1 + \ddot{\Psi}_1^\ast \Psi_1 + 4 \Psi_2^\ast \Psi_2+Q \gamma_0)
 ,\quad
\gamma_1=-2 i e(\dot{\Psi}_1\Psi^\ast_1  - \Psi_1 \dot{\Psi}^\ast_1)\ ,
\end{split}
\label{abcs}
\end{equation}
where $\alpha_0$ and $\gamma_0$ in the right hand side are given by the RNAdS$_4$ background.

Let us now compute the changes in the mass and charge caused by the superradiant scattering.
The Brown-York boundary stress tensor~\cite{Brown:1992br,Balasubramanian:1999re} is given by
\begin{equation}
 T_{ij} = \frac{1}{\kappa^2}(-K_{ij}+Kh_{ij}-2h_{ij}+G_{ij})-\Psi\Psi^\ast h_{ij} \ ,
\label{surfaceT}
\end{equation}
where $G_{ij}$ is the Einstein tensor with respect to the induced metric $h_{ij}$.
The last term of Eq.(\ref{surfaceT}) comes from the boundary term in the action of the scalar field~(\ref{fullaction}).
Let $\Sigma$ be a two-dimensional spacelike surface in the cutoff three-dimensional hypersurface at $z=\epsilon$.
The energy of the spacetime is given by 
\begin{equation}
 \mathcal{M}V = \int_{\Sigma} d^2 x  \sqrt{\sigma}  T_{ij} k^i n^j \ ,
\end{equation}
where $\sigma_{ab}$ is the induced metric of $\Sigma$, $n^i$ is the unit normal to $\Sigma$ and $k^i$ is the timelike Killing vector.
Taking $\Sigma$ as the constant-$t$ time slice, we find $\mathcal{M}=\int dxdy T_{tt}/\epsilon$. Computing the 
Brown-York boundary stress tensor by using the perturbed metric, we obtain the energy density as 
\begin{equation}
 \mathcal{M} = \frac{2M}{\kappa^2} + \beta_0 + \Psi_1 \Psi_2^\ast + \Psi_2 \Psi_1^\ast\ .
\end{equation}
From Eq.(\ref{abcs}), time derivative of the energy density is given by
\begin{equation}
 \dot{\mathcal{M}}=-\Psi_2^\ast \dot{\Psi}_1  - \Psi_2 \dot{\Psi}_1^\ast  \ ,
\end{equation}
which is precisely Eq.(\ref{dotM0}). We can also compute the charge density as 
\begin{equation}
 \mathcal{Q} = \frac{1}{2\kappa^2 V} \int dx dy F_{tz} = \frac{Q}{2\kappa^2} + \frac{\gamma_0}{2}\ .
\end{equation}
Its time derivative is 
\begin{equation}
 \dot{\mathcal{Q}} =-ie (\Psi_1^\ast \Psi_2 - \Psi_1 \Psi_2^\ast )\ ,
\end{equation}
which is nothing but Eq.(\ref{dotQ0}). 

\section{Quasinormal modes and instability of the RNAdS$_4$}
\label{sec:qnm}

In this appendix, we review the quasinormal modes and instability for the charged scalar field in the RNAdS$_4$ in relation to our numerical calculations in the main text. 

We take the Klein-Gordon equation \eqref{SchEq} and solve it with the vanishing source boundary condition $\psi(z) \to 0$ in the AdS boundary $(z \to 0)$ as well as the ingoing wave boundary condition \eqref{ingoing} on the horizon $z=z_+$. Then, regular solutions can be obtained for complex $\omega$, corresponding to the frequencies of the quasinormal modes. We numerically compute the frequencies by using pseudospectral methods with a Chebyshev grid. Note that $\Phi_H=Q z_+$ for the RNAdS$_4$, and we show results in units where $z_+=1$.

\begin{figure}[t]
  \begin{minipage}[t]{0.5\linewidth}
    \centering
    \includegraphics[width=7cm]{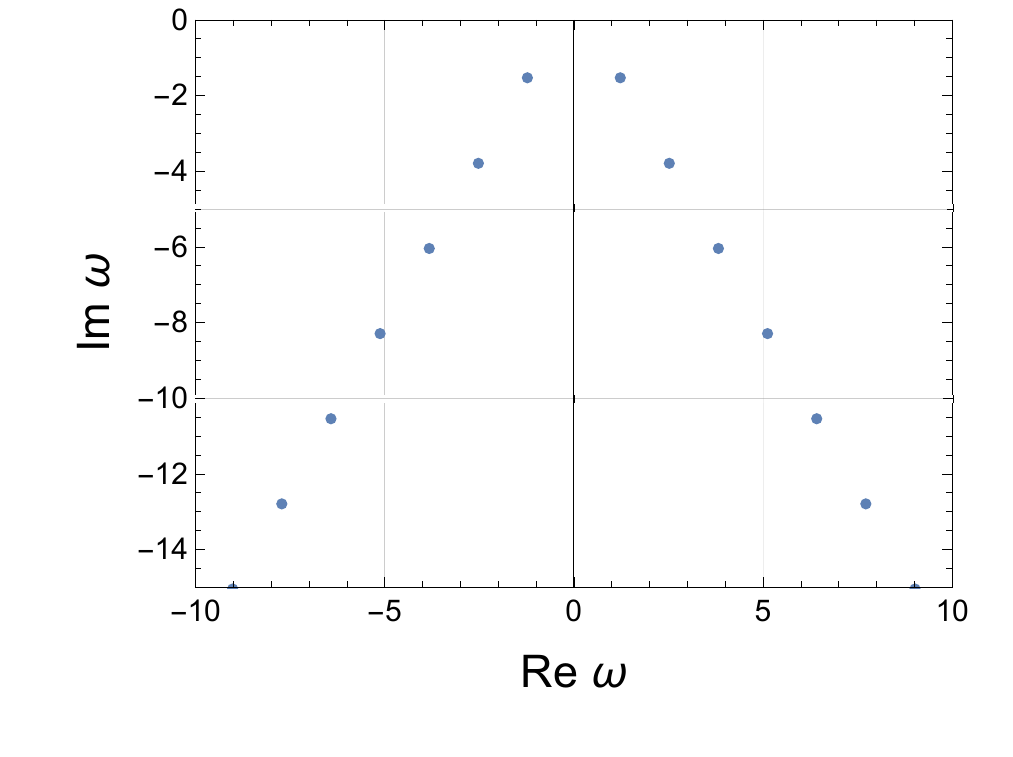}
  \end{minipage}
  \begin{minipage}[t]{0.5\linewidth}
    \centering
    \includegraphics[width=7cm]{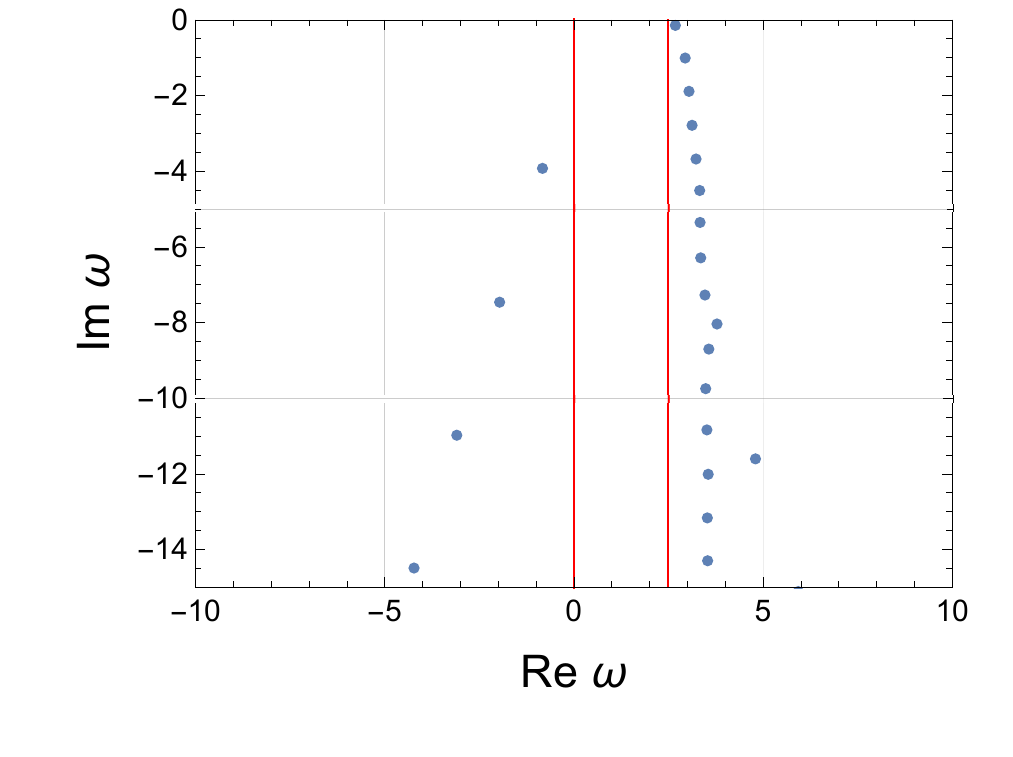}
  \end{minipage}
  \caption{Location of the quasinormal mode frequencies for $Q=0$ (left) and $Q=2.5$ (right) when $e=1$. The red vertical lines in the right panel show the boundaries of the superradiant condition $0 < \omega < e \Phi_H$.}
   \label{fig:qnm_w}
\end{figure}
Figure~\ref{fig:qnm_w} is the distribution of the quasinormal mode frequencies for $Q=0$ (left panel) and $Q=2.5$ (right panel) when $e=1$. In our convention, a frequency with a negative imaginary part corresponds to a stable mode decaying in time as $e^{(\mathrm{Im}\,\omega ) t}$. In the right panel, the behavior can be seen that an infinite number of modes comes in from the negative infinity of $\mathrm{Im}\,\omega$ when $Q\neq 0$. No stable modes ($\mathrm{Im}\,\omega < 0$) are in the range $0 < \mathrm{Re}\,\omega < e\Phi_H$. This is quite natural because it is counterintuitive if we have stable quasinormal modes that extract energy by superradiance.

\begin{figure}[t]
  \begin{minipage}[t]{0.5\linewidth}
    \centering
    \includegraphics[width=7cm]{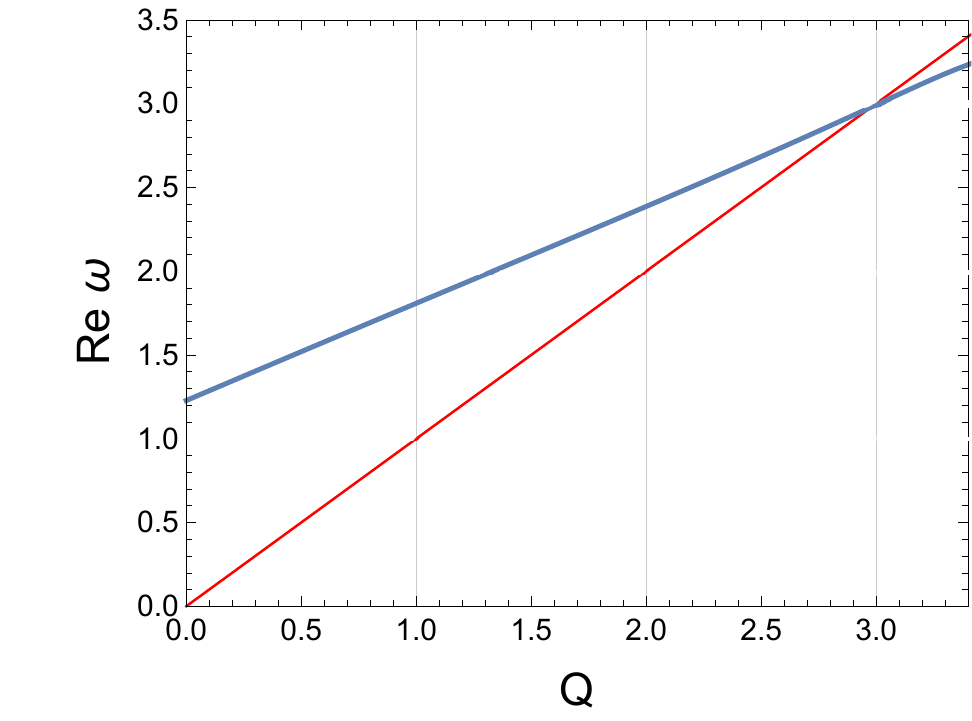}
  \end{minipage}
  \begin{minipage}[t]{0.5\linewidth}
    \centering
    \includegraphics[width=7cm]{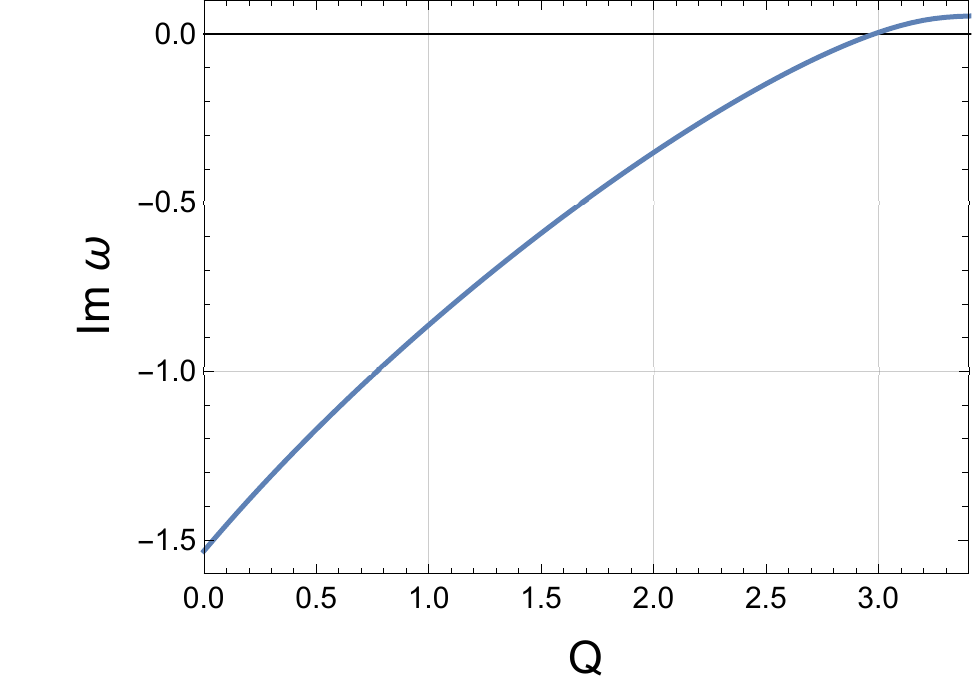}
  \end{minipage}
  \caption{The $Q$ dependence of the real and imaginary parts of the mode with the largest $\mathrm{Im}\,\omega$. The red line in the left panes shows the upper boundary of the superradiant condition $\omega=e\Phi_H$. In the right panel, the line crosses $\mathrm{Im}\,\omega=0$ at $Q=2.981$.}
    \label{fig:qnm_e0}
\end{figure}
In this system, $\mathrm{Im}\,\omega$ can become positive as $Q$ is increased, resulting in instability. In Fig.~\ref{fig:qnm_e0}, we show the behavior of the mode with the largest $\mathrm{Im}\,\omega$. The mode approaches the real axis as $Q$ is increased, and it crosses the axis at $Q=2.981$, where $\mathrm{Re}\,\omega=e\Phi_H$. This point is called the onset of instability. Beyond the onset, the unstable mode exists in $0 < \mathrm{Re}\,\omega < e\Phi_H$ with $\mathrm{Im}\,\omega > 0$.\footnote{For a small RNAdS$_4$, it was shown that $\mathrm{Im}\,\omega > 0$ if $0 < \mathrm{Re}\,\omega < e\Phi_H$ \cite{Uchikata:2011zz}. Our numerical results indicate that this condition is satisfied also in large black holes.} The growth of the peak in $\dot{\mathcal{M}}$ is correlated with the approach of the mode to the real axis of $\omega$, although the location of the peak in $\dot{\mathcal{M}}$ is not the same as $\mathrm{Re}\,\omega$ of this mode.

\bibliography{SRAdS}

\end{document}